\documentclass[sigconf,screen]{acmart}

\copyrightyear{2024}
\acmYear{2024}
\setcopyright{acmlicensed}
\acmConference[MET '24]{Proceedings of the 9th ACM International Workshop on Metamorphic Testing}{September 17, 2024}{Vienna, Austria}
\acmBooktitle{Proceedings of the 9th ACM International Workshop on Metamorphic Testing (MET '24), September 17, 2024, Vienna, Austria}
\acmDOI{10.1145/3679006.3685071}
\acmISBN{979-8-4007-1117-6/24/09}

\usepackage{microtype}
\usepackage{xcolor}
\usepackage{algorithm}
\usepackage{algpseudocode}
\usepackage{siunitx} %
\usepackage{bm} %
\usepackage{subcaption}

\title{Evaluating Human Trajectory Prediction with Metamorphic Testing}

\author{Helge Spieker}
\orcid{0000-0003-2494-4279}
\affiliation{%
  \institution{Simula Research Laboratory}
  \city{Oslo}
  \country{Norway}
}
\email{helge@simula.no}

\author{Nassim Belmecheri}
\orcid{0000-0003-3436-0154}
\affiliation{%
  \institution{Simula Research Laboratory}
  \city{Oslo}
  \country{Norway}
}
\email{nassim@simula.no}

\author{Arnaud Gotlieb}
\orcid{0000-0002-8980-7585}
\affiliation{%
  \institution{Simula Research Laboratory}
  \city{Oslo}
  \country{Norway}
}
\email{arnaud@simula.no}

\author{Nadjib Lazaar}
\orcid{0000-0003-2524-9462}
\affiliation{%
  \institution{LIRMM, University of Montpellier, CNRS}
  \city{Montpellier}
  \country{France}
}
\email{nadjib.lazaar@lirmm.fr}

\begin{document}

\begin{abstract}
    The prediction of human trajectories is important for planning in autonomous systems that act in the real world, e.g. automated driving or mobile robots. Human trajectory prediction is a noisy process, and no prediction does precisely match any future trajectory. It is therefore approached as a stochastic problem, where the goal is to minimise the error between the true and the predicted trajectory.
    In this work, we explore the application of metamorphic testing for human trajectory prediction.
    Metamorphic testing is designed to handle unclear or missing test oracles. It is well-designed for human trajectory prediction, where there is no clear criterion of correct or incorrect human behaviour.
    Metamorphic relations rely on transformations over source test cases and exploit invariants. A setting well-designed for human trajectory prediction where there are many symmetries of expected human behaviour under variations of the input, e.g. mirroring and rescaling of the input data.
    We discuss how metamorphic testing can be applied to stochastic human trajectory prediction and introduce the Wasserstein Violation Criterion to statistically assess whether a follow-up test case violates a label-preserving metamorphic relation.
\end{abstract}



\maketitle

\section{Introduction}

Human trajectory prediction (HTP) is the task of predicting the future paths that individual humans may take based on the trajectories of their past movements and environment.
It is a key component in many autonomous systems that must be aware of its environment, one being automated driving~\cite{Levinson2011}.
Here, a specific focus is on predicting future trajectories of other traffic participants, like pedestrians or cyclists.
This group is commonly referred to as Vulnerable Road Users (VRUs) and due to their difference in movement patterns to vehicles, predicting their trajectories is a separate task.

Human trajectory prediction is an active research area, and current methods have achieved strong results~\cite{li2022graph,xu2022groupnet,Bae_2022_CVPR,duan2022complementary,Shi_2021_CVPR,Dendorfer_2021_ICCV,Mohamed_2020_CVPR,mangalam2020not}.
Despite these recent advancements, ensuring the robustness, accuracy, and reliability of these prediction models is still a challenge. 
It is crucial to rigorously test these models to identify potential flaws, enhance their performance, and ensure their safe practical application~\cite{Uhlemann2024}.
Some work exists in the domains of adversarial testing~\cite{Zhang_2022_CVPR,cao2022advdo,pmlr-v205-cao23a,Zheng_2023_WACV,pmlr-v211-tan23a,jiao2022semi} and verification~\cite{Zhang_2023_ICCV} with a focus on creating specific failure cases.

Given that HTP models are machine learning systems and operate stochastically, testing does not only serve the detection of bugs, but also to measure the model performance.
While the HTP test data contains a ground-truth future trajectory, many alternative trajectories would have been similarly realistic and a broader evaluation scheme besides the distance to one ground-truth trajectory would give more information about robustness of the method~\cite{Mohamed2022}.

Applying metamorphic testing to human trajectory prediction addresses the complexities and non-determinism of these models, where traditional testing is hindered by the lack of precise ground-truth data and the challenge of obtaining accurate datasets. 
Metamorphic testing enhances robustness by identifying edge cases and subtle errors by validating metamorphic relations, which are properties the output should maintain under input transformations.
We envision the application of MT for human trajectory prediction in the sense of traditional testing, but also to expand the evaluation setting by providing a more diverse view of the robustness of the models under input transformations without requiring additional involvement in data collection and labelling.

In this work, we describe the problem setting of metamorphic testing for human trajectory prediction using a novel violation criterion to identify failed follow-up test cases.
We perform an illustrative experimental evaluation on an exemplary popular trajectory prediction system, namely YNet~\cite{Mangalam_2021_ICCV} on the popular Stanford Drone Dataset (SDD)~\cite{robicquet2016learning}.

\section{Background}

\subsection{Metamorphic Testing}
\label{sec:mt}
Some programs are considered as being {\it non-testable} because it is not possible to define complete and correct oracles for them \cite{Weyuker82}. Such non-testable programs include supervised machine learning models which generalise their prediction after being trained on a set of labelled instances \cite{Zhang2020}. The exact behaviour of these models largely depends on the datasets used to train them, and their predictions are usually marred by uncertainties. Metamorphic Testing (MT) is a software test input generation method which copes with non-testable programs by leveraging user-defined properties of the system, called \emph{metamorphic relations} and generating \emph{follow-up test cases} by using these relations \cite{Chen1998, Chen2018}. 
\begin{definition}[Metamorphic Relation (MR)]
Let $P$ be a program under test, $x$ and $y$ two test inputs for $P$, then a MR for $P$ is expressed as a relation $\forall x, \forall y, r_i(x,y) \implies r_o(P(x), P(y))$ where $P(x)$ (resp. $P(y)$) denotes the execution of $P$ on $x$ (resp. $y$). 
\end{definition}
It's worth noting that MRs are necessary (but not sufficient) properties to ensure the correctness of $P$ w.r.t. its specification. Formally speaking, $r_i(x,y) \wedge \neg r_o(P(x),P(y)) \implies \neg correct(P)$. MRs are convenient properties for generating follow-up test cases. 
\begin{definition}[Follow-up test cases]
Let $r_i(x,y) \implies r_o(P(x), P(y))$ be an MR for $P$, then if there exists a transformation $f$ (possibly non-deterministic) such that $y=f(x)$, then it becomes possible to generate a sequence of {\it follow-up test cases} from $x$, namely $<x,t(x), t(t(x)), ...>$ which all have to fulfil the MR for $P$. Thus, the transformation $t$ is convenient to generate follow-up test cases.
\end{definition}
As noted in Segura's survey~\cite{Segura2016}, many MRs can be identified for testing a program. A key difficulty in MT is thus to find MRs and determine which ones have the greatest fault-revealing capabilities. 
MT has been extensively used to test trained ML models including simple classifiers \cite{Murphy2008,Xie2011}, deep learning models~\cite{Ding2017a}, machine translation \cite{sun2018metamorphic}, object detection and classification~\cite{Spieker2020}. MT has received considerable attention in the automated driving field~\cite{Deng2022}, but to the best of our knowledge, it has not yet been applied to test human trajectory prediction models.

\subsection{Human Trajectory Prediction}

Recent research explores the prediction of human trajectories in different contexts. Multiple methods can be categorised based on the multimodality in predicting, the input signals fed to the prediction model, and the type of the output provided by the model.

Human trajectory prediction models have multiple modalities. Unimodal models assume that there is a single probable outcome or future path, multiple methods have been proposed such as Social
Forces \cite{helbing1995social}, Social LSTM \cite{Alahi2016SocialLH}. On the other hand, multimodal models consider multiple possible outcomes or future paths. This type of prediction acknowledges the inherent uncertainty in forecasting and provides several potential trajectories or trends based on the given data or input signals. Some multimodal models are based on generative aspects such as DESIRE~\cite{lee2017desire}, Trajectron++~\cite{salzmann2020trajectron++} and Introvert~\cite{shafiee2021introvert} where the idea is to generate the stochastic outcome in future predictions through a learned latent variable with a defined prior
distribution. Others, e.g. \cite{liang2020simaug,mangalam2020disentangling,zhao2021tnt}, are based on spatial probability estimates, where the multimodality is obtained through estimated probability maps.

Depending on the prediction model, multiple input signals can be expected, such as the human pose and gaze of other pedestrians in the scene. 
These signals can reveal the immediate intentions of the individual and the potential interactions, that can influence the trajectory of the individual. RGB scene images can offer a comprehensive view of the environment. Scene semantic representations and location data can provide context, thus enhancing the accuracy of the prediction.

Current trajectory prediction models are multimodal, meaning they take into account more than just the past motion of objects. They incorporate additional information, such as environmental maps. Furthermore, these models generate a stochastic output, representing multiple potential human trajectories, projecting the inherent uncertainty and variability in human behaviour and movement \cite{zhang2023pedestrian}.

\begin{definition}{Multimodal Human Trajectory Prediction}
Given the historical information, the objective of the model is to predict the distribution of a human's trajectory for $T$ future timesteps. The model learns the parameters $\theta$ of the probability $P_{\theta}(Y|X, M)$, where $X$ represents trajectory history, $M$ represents the map or environment information, and $Y$ represents the predicted trajectory.

For a given agent $i$, the model predicts the future positions in the next $T$ timesteps from the current time $t$, defined as $Y_i = (Y_{t+1}^i, Y_{t+2}^i, ..., Y_{t+T}^i)$.

The model's input is the historical information in the past $n$ timesteps, denoted as $X_i = (X_{t-n+1}^i, X_{t-n+2}^i, ..., X_t^i)$.

\end{definition}

\section{Related Work}
Forecasting the trajectory of pedestrians based on their past movements is important to design safe automated driving systems. Previous work has addressed the challenge of verifying the robustness of HTP models by considering adversarial attacks~\cite{Zhang_2022_CVPR,cao2022advdo,pmlr-v205-cao23a,Zheng_2023_WACV,pmlr-v211-tan23a,jiao2022semi}. However, many of these works have just translated adversarial attacks proposed in the context of image classification and object detection tasks without taking into account the peculiarities of HTP model robustness verification. Recently, by using probably approximately correct (PAC) learning and formalising the notion of HTP robustness, \citeauthor{Zhang_2023_ICCV} have proposed in \cite{Zhang_2023_ICCV} a rich framework to verify the robustness of pedestrian trajectory prediction models. Using ablation studies, \citeauthor{Uhlemann2024} have proposed to evaluate HTP model safety in the context of automated driving~\cite{Uhlemann2024}. These approaches are interesting, but they do not rely on systematic testing methodologies, which can detect faults consistently. 

To the best of our knowledge, MT has not yet been used for testing HTP models, but approaching the verification of stochastic systems with MT is not new \cite{Chen2018}. Introduced by \citeauthor{guderlei2007statistical}, statistical MT replaces traditional violation criteria, i.e., the detection of violated MR, with hypothesis testing \cite{guderlei2007statistical}. Used for testing statistical optimization algorithms, e.g., simulated annealing, statistical MT reveals itself interesting but also dependent on the problem to be solved regarding its performance \cite{Yoo2010}. We believe that this approach, i.e., statistical MT, is relevant for testing HTP models and explore its usefulness in this paper.

\section{Metamorphic Testing of Human Trajectory Prediction}

We present an MT method for human trajectory prediction (HTP), designed for handling stochastic prediction outputs. 
Current HTP models expect as input the previous trajectory of the human plus additional information. 
We consider HTP models where the additional information is a visualisation of the environment from birds-eye view, e.g. from a drone recording.
This is a common setup in the HTP literature~\cite{mangalam2020not,Mangalam_2021_ICCV,luo2023gsgformer,robicquet2016learning}.
To be useful for HTP, the image input is segmented to identify the different regions in the scene, which give an indication of which areas are walkable and which are more likely to be used.

\begin{figure}
    \centering
    \begin{subfigure}[t]{0.48\columnwidth}
        \centering
        \includegraphics[width=\textwidth]{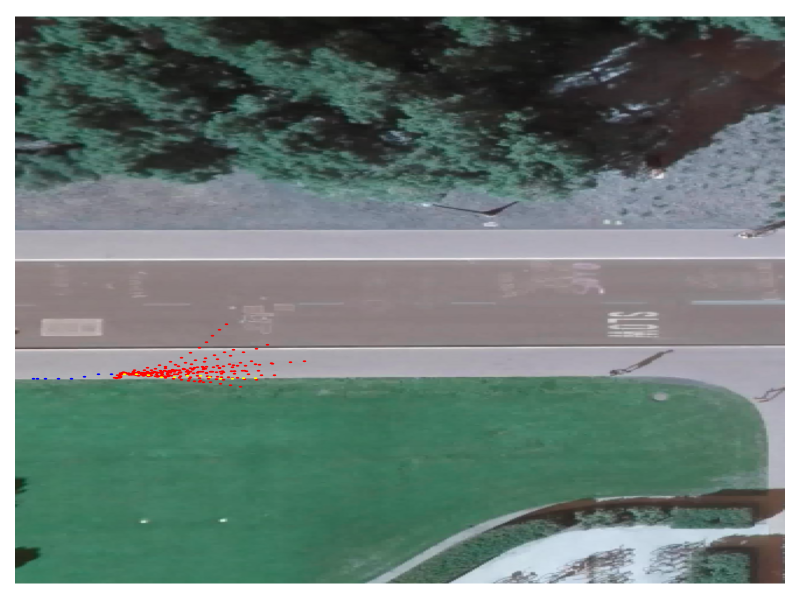}
        \caption{Inputs and Outputs. Blue is the past history, red is the predicted trajectories, yellow is the ground-truth trajectory.}
        \label{fig:input_output}
    \end{subfigure}
    \hfill
    \begin{subfigure}[t]{0.48\columnwidth}
        \centering
        \includegraphics[width=\textwidth]{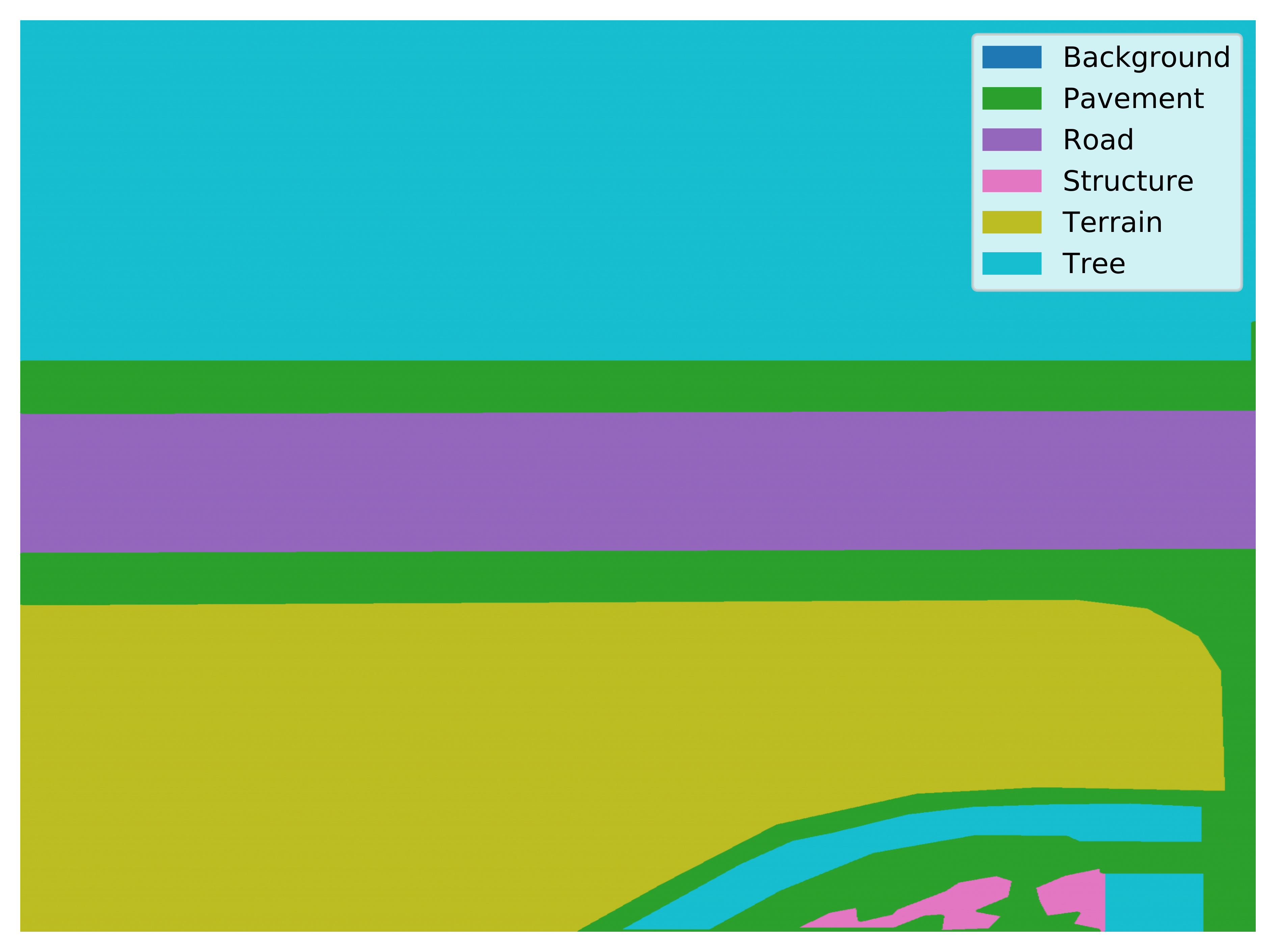}
        \caption{Image Segmentation Output. Six classes can be annotated.}
        \label{fig:segmentation}
    \end{subfigure}
    \caption{Inputs and Outputs of Human Trajectory Prediction. Data shows the little\_1 scene from the Stanford Drone Dataset~\cite{robicquet2016learning}.}
    \label{fig:htp_data}
\end{figure}

Both the historical trajectory and the environment can be modified by the metamorphic relations.
The historical trajectory can be manipulated directly, as it is a sequence of $(x,y)$ tuples.
Manipulating the image input directly is more difficult to do automatically and runs the risk of introducing unrealistic artefacts that hinder the testing process, even with modern generative ML models.
For this reason, we manipulate the input on the level of the segmented image, i.e. after the first input processing step.
Figure~\ref{fig:htp_data} visualises the inputs and outputs of a HTP model.
The left side shows the original RGB image, the input trajectory (blue) and a set of sampled output trajectories (red).
The right side shows the corresponding segmentation map of the RGB image with colour-coded areas. In this example, five different area types plus a background class are distinguished, which is common in the literature~\cite{Mangalam_2021_ICCV,luo2023gsgformer}, but other class structures are possible.

In the following, we will first introduce the metamorphic relations and the violation criterion for follow-up test cases before we bring everything together into the overall metamorphic testing process for HTP.

\subsection{Metamorphic Relations}

We introduce a set of metamorphic relations (MRs) to transform source test cases into follow-up test cases. A test case in HTP is the combination of segmentation map and input trajectory of past motion history.
The MRs considered in this study are mirroring the input in the horizontal or vertical axis and changing the rescaling factor of the segmentation map, i.e. resizing the map and trajectory.
All of these MRs are revertible, i.e. the source test case can be reconstructed from the follow-up test case, and label-preserving, i.e. if available ground-truth labels were transformed similarly they could be evaluated. 
For our violation criterion, ground-truth labels are not required.

\begin{definition}[MR1: Mirroring]
    The input is mirrored along the horizontal or vertical axis of the segmentation map.

Mirroring is a basic transformation, and the HTP model should be robust against these transformations.
Mirroring might lead to corruption when applied to the original image, for example, in Figure~\ref{fig:input_output} the label ``Slow'' on the street would be unreadable.
The segmentation map (Figure~\ref{fig:segmentation}) does not carry this level of detail and is not corrupted by the mirroring operation.
\end{definition}

\begin{definition}[MR2: Rescale]
    The rescaling factor of the original image is modified.

The intuition for this MR is that the original input images are resized before they are processed. 
However, the exact input size is not fixed and can be varied.
The distribution of trajectories should be consistent, independent of the size of the input image.
\end{definition}

The MRs presented here are not exhaustive, but they were selected for being intuitive and label-preserving, allowing us the evaluation of the Violation Criterion, that we will introduce next.

\subsection{Wasserstein Violation Criterion}

As part of the testing process, we must validate whether the result for the follow-up test case violates the MR.
This validation is performed through a violation criterion.
In many MRs, the violation criterion is a basic comparison, e.g. a violation occurs if the result of the follow-up test case is $\{=,\neq,\leq,\geq,<,>\}$ than the source test case result.
However, the HTP model is a stochastic system and returns a distribution of future trajectories.
Here, a basic comparison itself is not suitable, and we need a different violation criterion. This violation criterion must compare two distributions of trajectories and decide whether they are reasonably similar or whether there is a substantial difference, such that the underlying MR is violated.

We propose the \textit{Wasserstein Violation Criterion} (WVC) for the detection of faults in label-preserving MRs in HTP.
The WVC approaches the comparison of the two distributions as an optimal transport problem~\cite{peyre2019computational}, i.e. it determines the minimal cost to transform one distribution into the other.
Specifically, we compare the trajectory distribution using the \textit{Wasserstein} distance.
The Wasserstein distance is based on a matching between the sampled trajectories in each set, where the overall distance between the matches is minimal. Intuitively, it is described as the minimal cost to transform one probability distribution into the other, and also referred to as \textit{earth mover's distance}~\cite{rubner2000earth}, which visualises the optimal transport concept for two piles of earth that represent two distributions and should be compared by moving as little earth as possible.
For HTP, the Wasserstein distance is the minimum distance from the trajectories in one distribution to the other distribution, where each trajectory is assigned to exactly one other trajectory. The more similar the two distributions are, the smaller the Wasserstein distance.

Having the distance between outputs of the source and follow-up test case itself is not sufficient to decide whether the MR is violated.
Additionally, a threshold after which the distance becomes anomalous, and a violation is needed.
Since it is difficult to manually define a distance threshold from which a predicted trajectory should be counted as a violation and since it would have to be determined per scene and input trajectory, we adopt a stochastic approach instead:
We first sample multiple sets of trajectories for the source input, i.e. multiple predictions of the SUT, and calculate the mean and standard deviation of the pairwise Wasserstein distance between these sets.
For the follow-up test case, a violation then occurs if a z-test reports a significant ($\text{p-value} \leq threshold$) difference. The p-value threshold can be adjusted to the MR or kept at the common value of $0.05$. 

\subsection{Test Process}

Algorithm~\ref{alg:cap} outlines the metamorphic testing process for a single source and follow-up test case.
The process follows the general structure of metamorphic testing and its three phases:
First, the source test case is sampled and the system-under-test is executed with it.
In our case, to handle the non-determinism in the HTP model, we execute the SUT multiple times --- parametrized by the parameter $N$ --- and calculate the pairwise Wasserstein distances between the predictions and calculate statistics.
Afterwards, the test case is transformed according to the selected MR and executed once.
In the evaluation phase, the result of the follow-up test case is compared against each source test case execution and the z-test is calculated to detect potential violations.

\begin{algorithm}[t]
    \renewcommand{\algorithmicrequire}{\textbf{Input:}}
    \caption{Test Process Overview}\label{alg:cap}
    \begin{algorithmic}[1]
    \Require $HTP\text{: System-under-Test}$
    \State $SourceResults \gets \varnothing$
    \State $ViolationCounter \gets 0$
    \State
    \State $S \gets \text{Sample source test case}$ \Comment{Preparation Phase}
    \For{$i \gets 1$ to $N$}
        \State $r \gets HTP.predict(S)$
        \State $SourceResults \gets SourceResults \cup \{r\}$         
    \EndFor
    \State $D_{Src} \gets \text{PairwiseWassersteinDistances}(SourceResults)$
    \State $\langle \mu_{Src}, \sigma_{Src} \rangle \gets \text{CalculateVariationMeasures}(D_{Src})$

    \State
    \State $MR \gets \text{Select MR to apply}$
    \Comment{MT Phase}
    \State $FU \gets MR.transform(S)$
    \State $R_{FU} \gets HTP.predict(FU)$

    \State
    \For{$R_{S} \in SourceResults$} \Comment{Evaluation Phase}
        \State $r \gets SUT(S)$
        \State $R_{S}' \gets MR.transform(R_{S})$
        \State $D \gets \text{CalculateWassersteinDistance}(R_{FU}, R_{S}')$
        \State $PValue = ZTest(D, \mu_{Src}, \sigma_{Src})$
        \If{$PValue \leq 0.05$}
            \State $ViolationCounter = ViolationCounter + 1$
        \EndIf
    \EndFor
    \State \Return $ViolationCounter$
    \end{algorithmic}
\end{algorithm}

\section{Experiments}

\subsection{Experimental Setup}

We use the Stanford Drone Dataset (SDD)~\cite{robicquet2016learning}, which is widely used in the trajectory prediction literature~\cite{Mangalam_2021_ICCV,luo2023gsgformer}.
The dataset consists of 11,000 unique pedestrians in 8 top-down scenes around the Stanford University campus.
To avoid data leakage, we take the scenes from the test split of the dataset as in~\cite{Mangalam_2021_ICCV}.

Since we utilise the existing test set of SDD, we have ground-truth information available for our experiments.
We use this ground-truth information to calculate the standard trajectory prediction metrics ADE and FDE for the source and follow-up predictions. These metrics form a reference to interpret the effectiveness of the stochastic violation criterion and the general effect of the metamorphic transformations on prediction performance.

The system-under-test (SUT) is the YNet trajectory prediction model~\cite{Mangalam_2021_ICCV} using the publicly available, trained model weights\footnote{Online: \url{https://github.com/HarshayuGirase/Human-Path-Prediction}} and the experimental parameters.
We test two forecasting settings, following the experimental conditions from \citeauthor{Mangalam_2021_ICCV}~\cite{Mangalam_2021_ICCV}.
\textit{Short-term forecasting} has a $t_p = 3.2$ second past motion history, sampled at 2.5 FPS, and a prediction horizon of $t_f = 4.8$ seconds. 
\textit{Long-term forecasting} has a $t_p = 5$ second past motion history, sampled at 1 FPS, and a prediction horizon of $t_f = 30$ seconds. 
For both settings, YNet samples $K=20$ trajectories per prediction.

Per source test case, we sample $N=8$ sets of solutions to calculate the violation threshold and compare the follow-up test cases against it.
We report the violation rate, i.e. the percentage of prediction comparisons for which the distance exceeds the threshold, as the main metamorphic testing criterion.
We further calculate the average performance of the source and follow-up test in terms of average (ADE) and final displacement error (FDE), the standard evaluation metrics for HTP:
\begin{align}
    ADE &= \frac{1}{N \times T_p} \sum_{n \in N} \sum_{t \in T_p} \lVert \hat{p}_t^n - p_t^n \rVert_{2}\\
    FDE &= \frac{1}{N} \sum_{n \in N} \lVert \hat{p}_{T_p}^n - p_{T_p}^n \rVert_{2}
\end{align}
ADE is the average distance between the closest prediction and the ground-truth trajectory, FDE is the distance between the trajectory endpoints. The common evaluation setup is Best-of-N (BoN), which means, the smallest ADE and FDE are reported over N sampled trajectories, i.e. the $K=20$ in our case.
Since BoN evaluation does not consider the distribution of the trajectories besides the best one, we additionally calculate the mean ADE and FDE.
To identify MR violations, we apply a similar approach to the WVC and compare the ADE/FDE of the follow-up test case to the averaged results of all sampled source test cases via a z-test.
We denote the two sets of metrics as \textit{BoN-ADE}, \textit{BoN-FDE}, \textit{Mean-ADE}, and \textit{Mean-FDE}.
These four metrics require ground-truth information, which is generally not available in metamorphic testing. They are included in the experiments to evaluate the utility of the MRs and the WVC.

For the Rescale MR, we choose two different rescale values $0.2$ and $0.3$, that slightly deviate from the YNet default value of $0.25$.
These values are picked since they are close to the default value and should not introduce and too strong distribution shift for the model, but still cause the model input to be differently sized after all preprocessing steps.

Our implementation is based on the YNet codebase and uses POT (Python Optimal Transport) to calculate Wasserstein distances~\cite{flamary2021pot}.

\subsection{Results}

Table~\ref{tab:results_short} and Table~\ref{tab:results_long} summarise the results for each forecasting setting, separated per metamorphic relation.
We observe a close similarity in detected violations of the metamorphic relation for the proposed Wasserstein violation criterion, which does not need any ground-truth labels, and the ground-truth dependent Mean-ADE and Mean-FDE.
This similarity occurs in both forecasting settings.

There is further a strong difference in ADE/FDE values between BoN and Mean, especially in the short-term forecasting setting. Here, the forecasting horizon is smaller and the variation that the sampled trajectories can have is more limited than in the long-term forecasting setting.

\begin{table}[t]
    \centering
    \small
    \caption{Short-term forecasting: Violation rates (in \%) per metamorphic relation and compared to the labelled baselines. WVC: Wasserstein Violation Criterion; BoN: Best-of-N.}
    \label{tab:results_short}
    \begin{tabular}{l|rrrrr}
        \toprule
        MR & WVC & BoN-ADE & BoN-FDE & Mean-ADE & Mean-FDE \\\midrule
        Mirror-v & 61.0 & 27.0 & 27.8 & 65.2 & 64.4 \\
        Mirror-h & 61.9 & 26.1 & 25.1 & 64.6 & 63.5 \\
        Rescale  & 71.1 & 36.1 & 28.8 & 75.6 & 74.1 \\
        \bottomrule
    \end{tabular}

\end{table}

\begin{table}[t]
    \centering
    \small
    \caption{Long-term forecasting: Violation rates (in \%) per metamorphic relation and compared to the labelled baselines. WVC: Wasserstein Violation Criterion; BoN: Best-of-N.}
    \label{tab:results_long}
    \begin{tabular}{l|rrrrr}
        \toprule
        MR & WVC & BoN-ADE & BoN-FDE & Mean-ADE & Mean-FDE \\\midrule
        Mirror-v & 33.6 & 42.7 & 35.1 & 37.7 & 35.6 \\
        Mirror-h & 33.6 & 39.3 & 33.0 & 36.9 & 32.5 \\
        Rescale  & 44.1 & 40.8 & 33.2 & 48.0 & 43.3 \\
        \bottomrule
    \end{tabular}
\end{table}

We perform an additional experiment to investigate the agreement between the violations detected by WVC and the ADE/FDE-based criteria.
The experiment is approached as a binary classification problem, where the ADE/FDE-detected violations are considered the class labels and the WVC-detected violations are the predictions. We report accuracy, precision, and recall over multiple p-value thresholds to understand the sensitivity of the results, too.

\begin{figure}
    \centering
    \includegraphics[width=0.89\columnwidth]{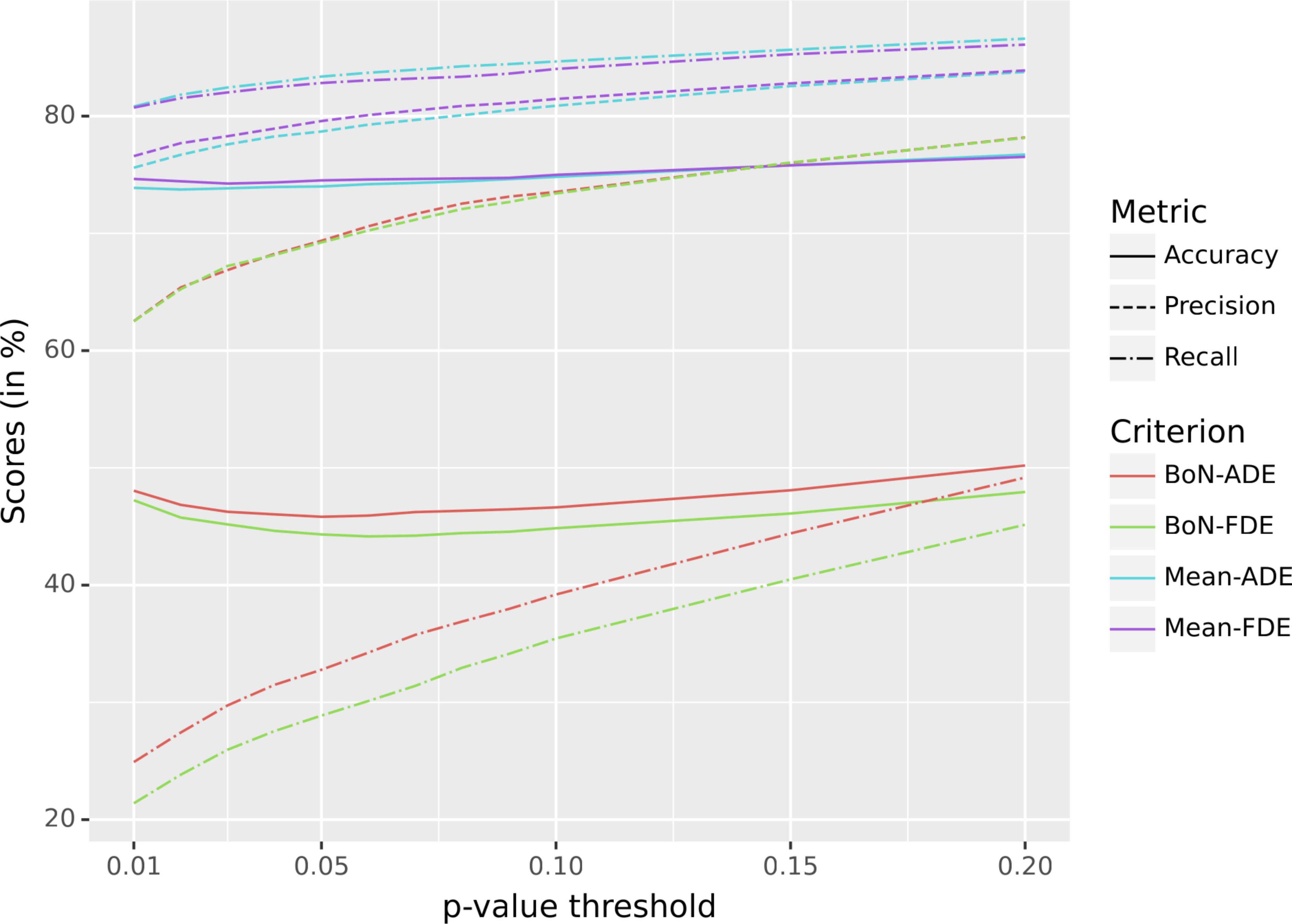}
    \vspace{-1.0em}
    \caption{Classification scores for short-term forecasting.}
    \label{fig:scores_shortterm}
\end{figure}

\begin{figure}
    \centering
    \includegraphics[width=0.89\columnwidth]{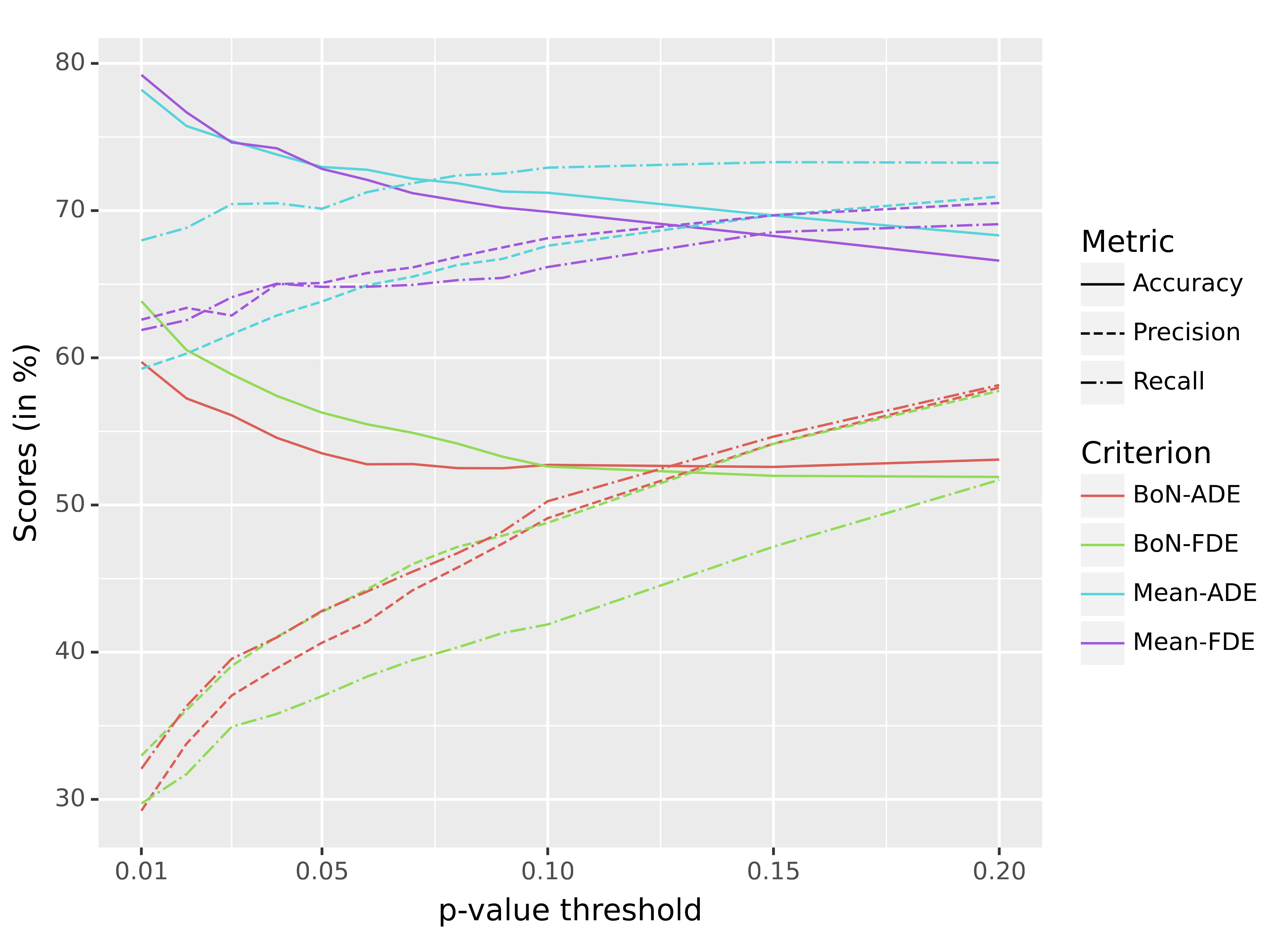}
    \vspace{-1.0em}
    \caption{Classification scores for long-term forecasting.}
    \label{fig:scores_longterm}
\end{figure}

The results are shown in Figures~\ref{fig:scores_shortterm} and \ref{fig:scores_longterm}.
They confirm for both settings that there is a substantial agreement between detected violations of WVC and Mean-ADE/FDE.
At the same time, they show that the p-value threshold is relevant to be adjusted per setting. In short-term forecasting, increasing the p-value threshold improves all metrics, whereas it has the opposite effect in long-term forecasting and decreases the overall classification quality.

\section{Conclusion and Future Work}

In this work, we have presented a metamorphic testing approach for human trajectory prediction (HTP).
Human trajectory prediction is a stochastic process, and we have adopted the MT methodology accordingly by introducing the statistical Wasserstein Violation Criterion, which identifies violations of label-preserving metamorphic relations through measuring the Wasserstein distances between predicted trajectory distributions and identifying statistically significant outliers.
Our experiments with a popular HTP model show that the proposed criterion is similarly effective as an alternative criterion that requires the annotated ground-truth data of the dataset to detect violations.

In future work, we will expand our evaluation over more metamorphic relations, trajectory prediction models, involving multiple pedestrians, and other base datasets in an attempt to enhance the existing evaluation setups by transformed and modified test scenarios.
Additionally, we will further consider the modelling of dedicated scenarios via custom segmentation maps and input trajectories for broader diversity in the scenarios. This should support the further automation of the metamorphic testing process~\cite{gotlieb2003automated}.

\section*{Acknowledgments}
This work is funded by the European Commission through the AI4CCAM project (Trustworthy AI for Connected, Cooperative Automated Mobility) under grant agreement No 101076911, the TAILOR project under agreement No 952215, and by the AutoCSP project of the Research Council of Norway, grant number 324674.

\bibliographystyle{ACM-Reference-Format}
\bibliography{refs}
\end{document}